\documentclass[draft,preprint,notoc]{JHEP3} 
\usepackage{epsfig}

\def\be{\begin{equation}}
\def\ee{\end{equation}}
\def\beq{\begin{eqnarray}}
\def\eeq{\end{eqnarray}}
\def\s{\sigma}

\def\G{\Gamma}

\def\hsr{hypergeometric series representations}

\def\ndim{NDIM}

\def\lra{\leftrightarrow}

\title{$\epsilon$-Expansion for non-planar double-boxes}

\author{Alfredo T. Suzuki \thanks{E-mail:suzuki@ift.unesp.br}\\
Universidade Estadual Paulista --- Instituto de F\'{\i}sica Te\'orica  \\
R. Pamplona, 145, S\~ao Paulo - SP, CEP 01405-900, Brazil}

\author{Alexandre G. M. Schmidt\thanks{E-mail:schmidt@fisica.ufpr.br} \\
Universidade Federal do Paran\'a --- Departamento de F\'{\i}sica  \\
Caixa Postal 19044, Curitiba - PR, CEP 81531-990, Brazil}

\date{\today}

\abstract{ We present calculations for non-planar double-box with
four massless/massive external/internal legs/propagators. The
results are expressed for arbitrary exponents of propagators and
dimension in terms of Lauricella's hypergeometric functions of
three variables and hypergeometric-like multiple series.}

\vspace{1cm} 

\received{...} \accepted{...}

\begin{document}

\section{Introduction}

Quantum field theories have come a long way since its inception
and today, all known interactions have been classed within its
model description for elementary particles in nature
\cite{glover-qcd}. Within this framework, studies on Feynman loop
integrals became even more compelling, challenging us with ever
increasing mathematical complexities associated with the
perturbative approach. Results for calculations of diagrams with
massive internal particles were presented by the authors
\cite{equiv} and others \cite{tramontano}; non-planar double-box
scalar and tensorial integrals were studied by Tausk \cite{tausk}
and Smirnov \cite{smirnov} using Mellin-Barnes technique ---
recently the triple box as well ---  while Gehrmann and Remiddi
 \cite{gehrmann} using the powerful differential equation method
calculated several 2-loop integrals. Bern and collaborators
studied dimensionally regularized one-loop pentagon integrals
 \cite{bern-penta} as well as Binoth {\it et al} \cite{binoth} (numerically) and
the authors (analytically and with arbitrary exponents of
propagators) tackle one-loop scalar hexagon integrals
\cite{n-point}.

These three methods are very powerful and interesting:
integration-by-parts relates a complicated integral to simpler
ones with some exponents of propagators raised to powers greater
than one. Differential equation method also relates a complicated
graph to simpler ones -- that means lesser number of loops or
simpler graphs with the same number of loops --, with the
exponents of propagators also changing when one uses this method.
Mellin-Barnes approach is based on an integral representation in
which the resulting integrals can be carried out summing up the
residues (either in the right or left complex plane). The latter
one therefore yields the two domains of validity available for the
hypergeometric functions connected by analytic continuation. These
in turn define the two distinct kinematical regions of interest
\cite{boos}.

An altogether different and elegant approach has been suggested
back in the middle of the 1980's by Halliday and Ricotta
\cite{halliday}, coined \ndim{}, where the key point is the
introduction of a negative dimensional space to work out the
integral. Their seminal ideia was reframed within the context of
solving systems of linear algebraic equations and from their
original work to our present understanding and experience in it we
deem \ndim{} to be a more compleat tool to handle complex Feynman
integrals, covariant and non-covariant alike, and far simpler in
its essence to implement, needing only to deal with systems of
linear algebraic equations of first degree. All kinematical
regions of interest come simultaneously defined and the bonus
by-product is that it defines even as yet unknown relationships of
analytic continuation among hypergeometric functions of several
variables.

Our aim in this work is to evaluate some loop integrals which were
not considered in the literature up to now using the \ndim{}
technique. They are double-box scalar integrals with six
propagators: in the simplest case, i.e., massless internal
particles and on-shell external legs the result \cite{tausk} is
well-known for a special case of exponents of propagators (all
equal to minus one); we fill the gap presenting the result for
arbitrary exponents of propagators and extend our knowledge
studying the same graph where four internal particles have
arbitrary mass $\mu$ and also studying the case where one of the
external legs is off-shell. These diagram computations become
important as progress in perturbative calculations for fundamental
interactions between particles are checked against the background
of our present experimental data increases their precision
measurements.

The outline for our paper is as follows: in section 2 we study
covariant four-point integrals with six propagators -- non-planar
double-box with four massive propagators -- the exact result of which
is written in terms of hypergeometric function of three variables. We
also perform numerical calculations in order to expand the result in
powers of $\epsilon$. Section 3 is concerned with massless non-planar
double-box with off-shell external legs and in section 4 we give
concluding remarks for our present work.

\section{Massive non-planar double-box}

To begin with consider the integral for the massive non-planar
double-box with six propagators, namely,
\be
{\cal I}_M\label{im} = \int d^D\! q\; d^D\! r\; (q^2\! -\! \mu^2)^i
[(q-p_3)^2\!-\!\mu^2]^j [(q+r)^2\!-\!\mu^2]^k [(q+r+p_2)^2\!-\!\mu^2]^l (r^2)^m
(r-p_4)^{2n}\,.
\ee

This is the scalar integral which arises in the computation of the
diagram depicted in figure 1, where the external legs represent
on-shell massless particles, i.e., $p_1^2=p_2^2=p_3^2=p_4^2=0$.

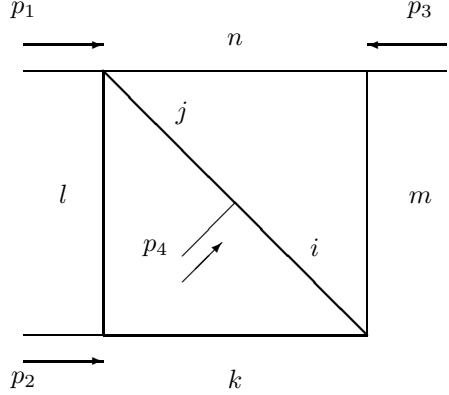
\begin{figure}
\begin{center}
\begin{picture}(600,200)(0,150)
\vspace{40mm} \thinlines \put(150,200){\line(1,0){30}}
{\thicklines \put(180,200){\line(1,0){100}}}
\put(150,190){\vector(1,0){30}} \put(150,300){\line(1,0){160}}
\put(150,310){\vector(1,0){30}} {\thicklines\thicklines
\put(180,200){\line(0,1){100}}} \put(280,300){\line(0,-1){100}}
{\thicklines\thicklines \put(280,200){\line(-1,1){100}}}
\put(230,250){\line(-1,-1){20}} \put(310,310){\vector(-1,0){30}}
\put(210,220){\vector(1,1){15}} {\small
\put(230,180){\makebox(0,0)[b]{$k$}}
\put(300,320){\makebox(0,0)[b]{$p_3$}}
\put(150,320){\makebox(0,0)[b]{$p_1$}}
\put(150,180){\makebox(0,0)[b]{$p_2$}}
\put(200,230){\makebox(0,0)[b]{$p_4$}}
\put(230,310){\makebox(0,0)[b]{$n$}}
\put(260,230){\makebox(0,0)[b]{$i$}}
\put(165,250){\makebox(0,0)[b]{$l$}}
\put(210,280){\makebox(0,0)[b]{$j$}}
\put(300,250){\makebox(0,0)[b]{$m$}}}
\end{picture}\caption{Scalar non-planar double-box with four massive propagators which are represented by
  thick lines. The labels in the internal lines represent the
  exponents of propagators. All external momenta are considered to be
  incoming, so $p_1+p_2+p_3+p_4=0$. }
\end{center}\end{figure}

The generating functional for {\em massless} on-shell non-planar
double-box\cite{pentabox} is given by,
\beq
G_0 &=& \int d^D\! q\;
d^D\! r\; \exp{\left\{\!-\alpha q^2\! -\!\beta (q-p_3)^2 \!-\!\gamma (q\!+\!r)^2
    \!-\!\theta (q\!+\!r\!+\!p_2)^2
    \!-\!\phi r^2 \right.} \nonumber\\
&& \left. -\omega (r-p_4)^2\right\} \\ &=&
\left(\frac{\pi^2}{\lambda}\right)^{D/2}
\exp{\left[-\frac{1}{\lambda}\left( \beta \gamma\omega s +
      \alpha\theta\omega t + \beta\theta\phi u \right)\right]},
\label{geradora-on} \eeq where we have defined
$\lambda'=\alpha+\beta+ \gamma+\theta$, $\lambda=\alpha\gamma+
\alpha\theta +\beta\gamma+ \beta\theta+ \lambda'(\phi+\omega)$ and
$s, t, u$ are the usual Mandelstam's variables, given by \be s=
2p_1\cdot p_2, \qquad t=2p_1\cdot p_3, \qquad u=2p_1\cdot p_4\,.
\ee

Since $p_i^2=0,\:\; (i=1,2,3,4)$, observe that $s+t+u=0$ follows from
the above equation.

Then, considering some of the internal particles to be {\em massive}, that
is, diagram of figure 1, one has the generating functional,
\be
G_M =\exp{(\lambda'\mu^2) }\,G_0,
\ee
with the {\em massive} sector factorized, and the following system of
algebraic equations,
\be
\label{sistema}
\left\{ \matrix{ X_2 + Y_{1234}+W_1 &=&
    i \cr X_{13} +Y_{5678}+W_2 &=& j \cr X_1+Y_{159}+W_3+Z_1 &=& k \cr
    X_{23}+Y_{26}+W_4+Z_{23}&=& l \cr X_3+Y_{379}+Z_2 &=& m \cr
    X_{12}+Y_{48}+Z_{13} &=& n \cr \Sigma X+\Sigma Y + \Sigma Z &=& -
    D/2 }\right.,
\ee where $W_j$ are the indices labelling the pure massive sector.
We use the shorthand notation,
$$
X_{abc} = X_a+X_b+X_c,
$$
and so on. In the last equation $\Sigma X = X_{123},\; \Sigma Y=
Y_{123456789},\; \Sigma Z= Z_{123}.$ Note that the total of sum
indices are 19 with 7 constraint equations, so that the result
will be a series of 12 indices. This 12-fold sum may be
constructed in various ways, in fact, $C^{12}_7$ ways. A huge
number of ways. The majority of them -- 30,972 -- yield vanishing
determinant for the algebraic system, so that the solution is
trivial in all of these cases. The remaining 29,416 yield
non-vanishing determinant and non-trivial solutions expressed as
\hsr{}. The several variables that identify these series belong to
a subset of the following set \be \left
\{1,\,\frac{t}{s},\,\frac{u}{t},\,\frac{s}{u},\,
            \frac{s}{t},\,\frac{t}{u},\,\frac{u}{s},\,
            \frac{t}{4\mu^2},\,\frac{u}{4\mu^2},\,\frac{s}{4\mu^2},\,
            \frac{4\mu^2}{t},\,\frac{4\mu^2}{u},\,\frac{4\mu^2}{s}\,\right\}\,.
\ee

The simplest of the \hsr{} for ${\cal I}_M$ is given by a {\em triple series},
\beq
{\cal I}_M &=& \pi^D (\mu^2)^\s\; \G \sum_{X_1,X_2,X_3=0}^\infty
\frac{X^{X_1}Y^{X_2}Z^{X_3}(-\s|X_{123})(-n|X_{12})(-m|X_{3})
  (-l|X_{23})}{X_1!X_2!X_3!(D/2|X_{123})(-i-j|X_{123})
  }\nonumber\\
&&\times \frac{(-i|X_2)(-j|X_{13})(-k|X_1) (-k-l-m-n-D/2|X_{123})}{(
  1/2-\s/2-m/2 -n/2|X_{123})(-\s/2-m/2-n/2|X_{123})},
\label{tripla}
\eeq
where $\s=i+j+k+l+m+n+D$, is the sum of exponents
and dimension, $(a|b)$ is the Pochhammer symbol,
$$
(a|b) \equiv (a)_b = \frac{\G(a+b)}{\G(a)},
$$
and
\be
\G = (-\s-m-n|m+n) (D/2|m+n)(-i-j|-m-n-D/2) (-k-l|-m-n-D/2),
\ee
where the subset of three variables are
$$
X=\frac{s}{4\mu^2}, \quad Y=\frac{t}{4\mu^2}, \quad
Z=\frac{u}{4\mu^2}.
$$

Observe that the final result, eq. (\ref{tripla}), has only a
three-fold series.  However, the expression provided by the solution
of system (\ref{sistema}) was a 12-fold series. It is very easy to
understand why this is so. Among the defining variables for the \hsr{}
here there are 9 whose variables are just unity and are summable
series. In other words, we were able to sum up nine of them using
Gauss' summation formula\cite{luke}.  Our strategy is therefore to
choose as many series as possible in which the individual sums can be
written as $_2F_1(a,\,b;\,c|1)$, then we plug them in a computer
program that do the job, i.e., sum up the series using Gauss'
summation formula.

Once we have evaluated this result, it is a straightforward exercise
to write down several other solutions which are connected by symmetry
in the diagram, namely, by exchanging the pairs $(i\leftrightarrow
k,\,s\leftrightarrow t)$, and $(j\leftrightarrow l,\,s\leftrightarrow
t)$


If we are interested in the primary integral where all the exponents
of propagators are equal to minus one, the above result reduces to
\beq
{\cal I}_M\label{4mass} &=& \pi^D (\mu^2)^{D-6} \G(\{-1\})
\sum_{X_1,X_2,X_3=0}^\infty
\frac{X^{X_1}Y^{X_2}Z^{X_3}(6-D|X_{123})(1|X_{12})(1|X_{3})
  (1|X_{23}) }{X_1!X_2!X_3!(D/2|X_{123})(2|X_{123}) }\nonumber\\
&&\times \frac{(1|X_1)(1|X_{13})(1|X_2)}{(9/2-D/2|X_{123})},
\eeq
where
\be
\G(\{-1\}) = \frac{\G(D/2-2)\G(D/2)\G(6-D)\G^2(4-D/2)}{\G(8-D)}\,.
\ee

The original system of linear equations defines a $ 7\times 19 $
rectangular matrix. From it we can draw $50,388$ square submatrices of
dimension $7\times 7$, of which $30,972$ yield vanishing determinant,
as already said before. Of course, among the 29,416 solvable solutions
that remain, \ndim{} provides other kind of series, such as 5-fold and
7-fold series, i.e., \hsr{} with five and seven variables respectively
(meaning seven and five summable series expressed as
$_2F_1(a,b;\,c|1)$ respectively).  And all of them have symmetries
among $s,t$ and $u$, namely,
\beq
&& (p_3\leftrightarrow p_4,\; j\lra n, \; i\lra m, \; t\lra u),\, \quad
(p_2\leftrightarrow p_3,\; l\lra n, \; k\lra m, \; t\lra s),\,\nonumber\\
&& (p_2\leftrightarrow p_4,\; j\lra l, \; i\lra k, \; s\lra u),\,
\label{simetrias} \eeq so, for each \hsr{} provided by \ndim{}
there are other two, also originated from the system of algebraic
equations, which represent the same integral and can be
transformed in the first using (\ref{simetrias}). This is the case
of (\ref{tripla}).

\begin{figure}
\begin{center}
\begin{picture}(600,200)(0,150)
\vspace{40mm} \thinlines \put(150,200){\line(1,0){130}}
\put(150,190){\vector(1,0){30}} \put(150,300){\line(1,0){160}}
\put(150,310){\vector(1,0){30}} \put(180,200){\line(0,1){100}}
\put(280,300){\line(0,-1){100}} \put(280,200){\line(-1,1){100}}
\put(230,250){\line(-1,-1){20}} \put(310,310){\vector(-1,0){30}}
\put(210,220){\vector(1,1){15}} {\small
\put(230,180){\makebox(0,0)[b]{$k$}}
\put(300,320){\makebox(0,0)[b]{$p_3$}}
\put(150,320){\makebox(0,0)[b]{$p_1$}}
\put(150,180){\makebox(0,0)[b]{$p_2$}}
\put(200,230){\makebox(0,0)[b]{$p_4$}}
\put(230,310){\makebox(0,0)[b]{$n$}}
\put(260,230){\makebox(0,0)[b]{$i$}}
\put(165,250){\makebox(0,0)[b]{$l$}}
\put(210,280){\makebox(0,0)[b]{$j$}}
\put(300,250){\makebox(0,0)[b]{$m$}}}
\end{picture}\caption{Scalar massless non-planar double-box with six propagators.}
\end{center}\end{figure}
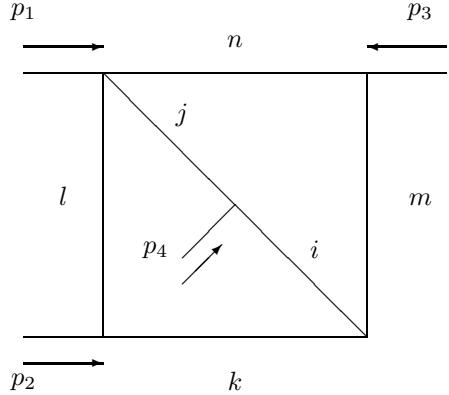

\subsection{Numerical calculation}

Expansion in $\epsilon$ for the integral (\ref{im}) can be obtained
numerically to all orders since our result (\ref{4mass}) is exact.
Hypergeometric series converge very fast and eq.(\ref{4mass}) can be
truncated after just few terms. We consider two examples, namely, $s=-1,
t=-2, s=-3, 4\mu^2=25$ and $s=2, t=-1, u=-1, \mu=4$.

\begin{center}
  Table
\end{center}
\begin{center}
\begin{tabular}{|l|c|c|c|}
\hline N & $a$ & $b$ & $c$ \\
\hline 1 & -0.15877787089947 & 0.1144504436314 & -0.4375363472289
\\ \hline 2 & -0.15905593258958 & 0.1141346540498 & -0.4382967223214 \\
\hline  3 & -0.15904157598309 & 0.1141585957166 & -0.4382477388046
\\ \hline 4 & -0.15904248795073 & 0.1141566914709 & -0.4382515309963  \\
\hline 5 &  -0.15904248795073 & 0.1141568521924 & -0.4382512066856
\\
\hline  6 & -0.15904242717753 & 0.1141568379462 & -0.4382512361932

\\ \hline 7 & -0.15904242673854 & 0.1141568392587 & -0.4382512333925 \\ \hline  8
& -0.15904242677727 & 0.1141568391341 & -0.4382512336666 \\
\hline 9
& -0.15904242677373 & 0.1141568391462 & -0.4382512336391 \\
\hline 10 & -0.15904242677406 & 0.1141568391450 & -0.4382512336419 \\
\hline 20 & -0.15904242677403 & 0.1141568391451 & -0.4382512336417 \\
\hline
\end{tabular}
\end{center}
N=Number of terms of each series and coefficients of
$\epsilon$-expansion for (\ref{im}). Case-I: $s=-1, t=-2, u=-3,
4\mu^2=25$. It is outside the physical region and therefore represents
a numerical sample. The result is in the form $\frac {a}{\epsilon} + b +
c\epsilon$, where $a,b,c$ are given in the table. It is possible to
obtain the $\epsilon$-expansion to all orders, since our result is
exact.

\vfill\eject

\begin{center}
  Table
\end{center}
\begin{center}
\begin{tabular}{|l|c|c|c|}
\hline N & $a$ & $b$ & $c$ \\
\hline 1 & -0.16666202713362544092 &
    0.11111642984629631817 &
   -0.4557060127212616592 \\
 \hline 2 & -0.16668058021298683828 & 0.11109516323349484374 &
 -0.4557569968657018264 \\ \hline  3 & -0.16668072454737026304 & 0.11109492182787342273 &
 -0.4557574905733665544
\\ \hline 4 & -0.16668072849754143706 & 0.11109491356240756367 & -0.4557575070366868277 \\
\hline 5 &  -0.16668072856344380806 & 0.11109491340169955725 &
-0.4557575073610663256
\\
\hline  6 & -0.16668072856499855487 & 0.11109491339744859266 &
-0.4557575073698739456
\\ \hline 7 & -0.16668072856503197638 & 0.11109491339734860742 & -0.4557575073700873608 \\ \hline  8
& -0.16668072856503277706 & 0.11109491339734602934 & -0.4557575073700930307\\
\hline 9
& -0.16668072856503279614 & 0.11109491339734596400 & -0.4557575073700931786 \\
\hline 10 & -0.16668072856503279661 & 0.11109491339734596228 & -0.4557575073700931826\\
\hline 20 & -0.16668072856503279662 & 0.11109491339734596224 & -0.4557575073700931827 \\
\hline
\end{tabular}
\end{center}
N=Number of terms of each series and coefficients of
$\epsilon$-expansion for (\ref{im}). Case-II: $s=2, t=-1, u=-1,
\mu=4$.  Observe that hypergeometric series converges very fast:
twenty terms for each of the three series provides 18 figures
precision. The result is in the form $\frac{a}{\epsilon} + b +
c\epsilon$, where $a,b,c$ are given in the table.

\section{Massless Double-box}

Our method can also be used to study Feynman diagrams where external
legs are off-shell\cite{probing}. Consider for instance the {\em massless}
non-planar double-box of figure 1. Let $p_i^2\neq 0$, ($i=1,2,3,4$),
that is, all external legs off-shell. The generating functional is
more complicated, namely,
\be
G_{4-OFF} = G_0
\exp{\left(-\beta\theta\omega\, p_1^2 - a_2\, p_2^2 - a_3\,p_3^2 -
    a_4\,p_4^2\right) },
\ee where $G_0$ is the generating functional (\ref{geradora-on})
for the on-shell massless diagram and \beq a_2 & = &
\alpha\gamma\theta+ \alpha\theta\phi+
\beta\gamma\theta +\gamma\theta\phi + \gamma\theta\omega \\
\nonumber a_3 & = & \alpha\beta\gamma +\alpha\beta\theta+
\alpha\beta\phi + \alpha\beta\omega +\beta\gamma\phi \\ \nonumber
a_4 & = & \alpha\gamma\omega+ \alpha\phi\omega + \beta\theta\phi+
\gamma\phi\omega+ \theta\phi\omega. \eeq

In the previous case we used the result $s+t+u=0$, however, for the
case at hand Mandelstam variables are,
\beq
s & = & (p_1+p_2)^2=(p_3+p_4)^2 \\ \nonumber
t & = & (p_1+p_3)^2=(p_2+p_4)^2 \\ \nonumber
u & = & (p_1+p_4)^2=(p_2+p_3)^2,
\eeq
and $s+t+u=p_1^2+p_2^2+p_3^2+p_4^2 \neq 0$.

In the next subsections we will study some particular cases. See Table

\begin{center}
  Table: Number of systems, solutions and type of results
\end{center}
\begin{center}
\begin{tabular}{|l|c|c|c|c|}
\hline Diagram & 4 Equal Masses & Massless (on) & Massless (I) &
Massless (II) \\ \hline System & $7\times 19$ & $ 7\times 15 $ &
$7\times 16 $ & $7\times 20$ \\ \hline Total number & 50,388 &
5,040 & 11,440 & 77,520 \\ \hline  Solutions & 29,416 & 2,916 &
4,632 & 34,994 \\ \hline Result & Triple Series & Double Series &
$F_A^{(3)}$ & 9-fold series \\ \hline
\end{tabular}
\end{center}

\subsection{One leg off-shell, $p_1^2\neq 0$}

There are two distinct cases to consider when one external leg is
off-shell: $p_1^2\neq 0$ and $p_j^2\neq 0$, for $j=2,\, 3,\,4$, since
the diagram is symmetric under the change $$
p_2 \lra p_3,\qquad p_2
\lra p_4, \qquad p_3 \lra p_4\,. $$
Observe that the vertex where $p_1$ is attached is quartic and all the
others are triple and can be interchanged leaving the diagram
unchanged.

Mandelstam variables must be rewritten as, \beq s&=&p_1^2
+2p_1\cdot p_2 \\\nonumber t&=&p_1^2 +2p_1\cdot p_3 \\\nonumber
u&=&p_1^2 +2p_1\cdot p_4, \eeq and $s+t+u=p_1^2=M_1^2$. So the
generating functional becomes slightly different,
\beq\label{geradora-off} G_{1-OFF} &=&
\left(\frac{\pi^2}{\lambda}\right)^{D/2}\!\!
\exp{\left[-\frac{1}{\lambda}\left( \beta \gamma\omega s +
      \alpha\theta\omega t + \beta\theta\phi u + \beta\theta\omega
      M_1^2 \right)\right]} \\ & =& G_0
\exp{\left(-\frac{\beta\theta\omega M_1^2}{\lambda}\right)},
\nonumber\eeq comparing (\ref{geradora-off}) with
(\ref{geradora-on}) we conclude that the original system gained
only one variable, i.e., the former system (double-box with 4 legs
on-shell) was $7\times 15$ and the present (double-box with 1 leg
off-shell) is $7\times 16$. The total number of solutions is now
11,440 being 6,808 trivial systems and we must deal with 4,632
possible ones.

The simplest \hsr{} for ${\cal I}_{1-off}$ are triple series, \be
{\cal I}_{1-off} = \pi^D f_1 \sum_{X_{1,2,3=0}}^\infty
\frac{P_1^{X_1}P_2^{X_2}P_3^{X_3}(-\s|X_{123})(-k|X_{1})(-i|X_{2})
  (-m|X_{3})}{X_1!X_2!X_3!(1+l-\s|X_{1})(1+j-\s|X_{2})(1+n-\s|X_{3})
  }\label{tripla-1off-p1},\ee where $ P_1 = s/M_1^2,\; P_2=t/M_1^2,\;
P_3=u/M_1^2,$ and following the usual approach for massless
diagrams in the \ndim{} context we have summed up 6 series. The
above triple series is a Lauricella's function of three
variables\cite{luke}, namely, \be {\cal I}_{1-off} = \pi^D f_1
F_A^{(3)}\left.\left[
    \matrix{ -\s; -k, -i, -m\cr 1+l-\s, 1+j-\s, 1+n-\s}\right|P_1,
  P_2, P_3\right] ,\ee where for convergence $|P_1|+|P_2|+|P_3|<1$. We
also define, \beq f_1 &=&
(M_1^2)^\s (\s+D/2|-2\s-D/2) (i+j+m+n+D|-m-n-D/2)\nonumber \\
&&\times (k+l+m+n+D|-k-l-D/2) (i+j+k+l+D|-i-j-D/2),\nonumber\\
&&\times (-j|\s)(-l|\s)(-n|\s), \eeq the symmetries of the diagram \be
i\lra k, \;\; j\lra l,\;\; s\lra t;\qquad\!  i\lra m, \;\; j\lra
n,\;\; s\lra u;\qquad\! k\lra m, \;\; l\lra n,\;\; t\lra u, \ee are
expressed in our final result.

\subsubsection{Special Case}

In the special case where $i=j=k=l=m=n=-1$ we have, \be {\cal
I}_{1-off} = \pi^D f_1(\{-1\}) F_A^{(3)}\left.\left[ \matrix{ 6-D;
      1, 1, 1\cr 6-D, 6-D, 6-D}\right|P_1, P_2, P_3\right],\ee with
\be f_1(\{-1\}) = (p_1^2)^{6-D} \frac{\G^3(D-5)\G^3(D/2-2)
  \G(6-D)}{\G^3(D-4) \G(3D/2-6)}, \ee which have a double pole in the
$D=4-2\epsilon$ limit.

If one were interested in writing the above result in terms of
more complicated functions, namely, logarithms, polylogarithms and
$S_{a,b}$ integrals, then the following integral representation
for $F_A^{(3)}$, is in order \beq F_A^{(3)} \left.\left[ \matrix{
\alpha;
      \beta, \beta', \beta'' \cr \gamma , \gamma',
      \gamma''}\right|P_1, P_2, P_3\right] &=& \frac{1}{\Gamma_A}
\int_0^1 dx_1 dx_2 dx_3 \frac{x_1^{\beta-1} x_2^{\beta'-1}
  x_3^{\beta''-1}(1-x_1)^{ \gamma-\beta-1} }{(1-x_1P_1 -x_2P_2 -x_3P_3)^\alpha} \nonumber\\
&&\times (1-x_2)^{ \gamma'-\beta'-1} (1-x_3)^{ \gamma''-\beta''-1},
\eeq where the integration is constrained to $x_1+x_2+x_3=1$ and
$$
\Gamma_A = \frac{\G(\beta)\G(\beta')
  \G(\beta'')\G(\gamma-\beta)\G(\gamma'-\beta')
  \G(\gamma''-\beta'')}{\G(\gamma)\G(\gamma') \G(\gamma'')}, $$
in the
present case one get, \beq F_A^{(3)} \left.\left[ \matrix{ 6-D; 1, 1,
      1 \cr 6-D , 6-D, 6-D}\right|P_1, P_2, P_3\right] &=&
\frac{1}{\Gamma_A} \int_0^1 dx_1 dx_2 dx_3 \frac{(1-x_1)^{ 4-D}
  }{(1-x_1P_1 -x_2P_2 -x_3P_3)^{6-D}}\nonumber\\
&&\times (1-x_2)^{4-D} (1-x_3)^{ 4-D}, \eeq then one take
$D=4-2\epsilon$ and uses Taylor expansion. However, to carry out
the integral of second derivatives of such integral representation
can not be an easy task. For this reason we claim hypergeometric
series representations are simpler than the ones in terms of
polylogarithms: fast convergence, compact expressions and analytic
continuation relations among (i.e. kinematical regions) them.

\subsection{One leg off-shell, $p_2^2\neq 0$}
Now we turn to the last case, the one where the external leg
attached to a quartic vertex is off-shell. The generating
functional is, \be\label{geradora-p2} G_{1-OFF} = G_0
\exp{\left[-\frac{ \alpha\gamma\theta+ \alpha\theta\phi+
      \beta\gamma\theta +\gamma\theta\phi +
      \gamma\theta\omega}{\lambda} p_2^2 \right] }, \ee we see
immediately that there will be four extra sums. Simply compare
(\ref{geradora-p2}) and (\ref{geradora-off}), the former has four
arguments (which properly expanded in Taylor series will produce the
referred extra sums) more than the latter.

The system of algebraic equations will be slightly different than
(\ref{sistema}), i.e., for the present case one have,
\be\label{sistema-p2} \left\{ \matrix{ X_2 + Y_{1234}+U_{12} &=& i \cr
    X_{13} +Y_{5678}+U_{3} &=& j \cr X_1+Y_{159}+Z_1+U_{1345} &=& k
    \cr X_{23}+Y_{26}+Z_{23}+ U_{12345} &=& l \cr X_3+Y_{379}+Z_2
    +U_{24} &=& m \cr X_{12}+Y_{48}+Z_{13} +U_{5} &=& n \cr \Sigma
    X+\Sigma Y + \Sigma Z+\Sigma U &=& - D/2 }\right., \ee note that
as the diagram does not have massive internal lines, the system
have not "variables" (sum indices) $W_j$. Indices $U_j$ are
concerned with Taylor expansion of,

$$\exp{\left[-\frac{ \alpha\gamma\theta+
      \alpha\theta\phi+ \beta\gamma\theta +\gamma\theta\phi +
      \gamma\theta\omega}{\lambda} p_2^2 \right] }. $$

Such system has in principle 77,520 ($C_{7}^{20}$) solutions.
Determinant vanish in 42,526 of them, i.e., we must search
hypergeometric type functions representing the original two-loop
integral in a total of 34,994, less than half of possible
solutions for the $7\times 20$ system. However, in this case such
series in far more complicated, since they are 13-fold ones. The
results of such analysis will be presented elsewhere.

\section{Conclusion}
We studied in this work several integrals pertaining to non-planar
double-box diagrams. Firstly we considered the case where four
internal particles have mass $\mu$, then write down the result in
terms of a triple hypergeometric series. The second part deals with
massless internal particles but off-shell external legs. We calculated
the generating functional for our negative-dimensional integrals and
then presented some particular cases of interest.

\acknowledgments{ ATS and AGMS gratefully acknowledge CNPq
(301374/2001-5) for partial and full financial
  support, respectively.}

\end{document}